# City Data Fusion: Sensor Data Fusion in the Internet of Things

Meisong Wang[**]   Charith Perera[+]   Prem Prakash Jayaraman[*++]   Miranda Zhang[**]
Peter Strazdins[**]
Rajiv Ranjan[*]

[**]Reasearch School of Computer Science, Australian National University, Canberra, Australia
[+]The Open University, Milton Keynes, UK
[*]Digital Productivity, CSIRO, Australia
[++]RMIT University, Melbourne, Australia

**ABSTRACT**

Internet of Things (IoT) has gained substantial attention recently and play a significant role in smart city application deployments. A number of such smart city applications depend on sensor fusion capabilities in the cloud from diverse data sources. We introduce the concept of IoT and present in detail ten different parameters that govern our sensor data fusion evaluation framework. We then evaluate the current state-of-the art in sensor data fusion against our sensor data fusion framework. Our main goal is to examine and survey different sensor data fusion research efforts based on our evaluation framework. The major open research issues related to sensor data fusion are also presented.

## 1. INTRODUCTION

During the past decade, the Internet of Things (IoT) has gained significant attention in academia as well as industry [56]. The main reason behind this is the capabilities that IoT promises to offer. It promises to create a smart world where all the objects around us are connected to the Internet and communicate with each other with minimum human intervention [68].

Even though IoT encompasses a number of ideas and concepts, it does not have a clear definition. However, Tan and Wang [45] have defined IoT in a fairly comprehensive manner as *"Things have identities and virtual personalities operating in smart spaces using intelligent interfaces to connect and communicate within social, environment, and user contexts [45, 58]*. Some other definitions are presented in [5]. The papers [5, 79, 58] have surveyed the definition of IoT in three different perspectives: things, the Internet and semantics.

IoT enables the vision *"from anytime, anyplace connectivity for anyone, we will now have the connectivity for anything [76]"*. Further expanding this idea, the European Union has defined the above vision as *"The IoT allows people and things to be connected Anytime, Anyplace, with Anything and Anyone, ideally using Any network and Any service [25]"*.

The term Internet of Things was firstly coined by Kevin Ashton [4] in a presentation in 1998. He has also mentioned *"The IoT has the potential to change the world, just as the Internet did. Maybe even more so. [68]"*. Then, MIT presented their IoT vision in 1999. Later, IoT was formally introduced by the International Telecommunication Union (ITU) by *ITU Internet report* in 2005 [76].

The rest of the paper is organised as follows. Section 2 provides an overview of sensor networks. Section 3, sensor data fusion is defined and techniques are discussed. We also outline the possible extensions to improve sensor data fusion. In Section 4, we highlight the importance of data fusion in smart city applications. In Section 5 presents the evaluation framework that we used to evaluate different research efforts. We survey various sensor data fusion efforts and its importance towards IoT in the Section 6. Final section concludes the survey by highlighting the survey results and research gaps.

## 2. SENSOR NETWORKS

Sensor networks are the major enabler of the IoT. A *sensor* can be defined as a device that detects or measures a physical phenomenon such as humidity, temperature, etc. A *sensor node* is a physical platform that hosts one or more sensors. Each sensor node has the capability to sense, communicate and process data. A typical *sensor network* [2] comprises two or more sensor nodes which communicate between each other using wired and wireless means. In sensor networks, sensors can be homogeneous or heterogeneous. Multiple sensor networks can be connected together through different mechanisms. One such approach is through the Internet.

Typically, sensor nodes are deployed in densely manner around the phenomenon which we want to sense [2]. These sensor nodes are low-cost and small in size, that enable large deployments. Sensor network is not a concept that emerged with the IoT. The concept of sensor network and related research existed long time before the IoT was defined. This can be clearly seen when we evaluate the literature in the field. However, with the emergence of the IoT has facilitated the mainstream adoption of sensor network as a major technology used to realise the IoT vision.

In recent times, another widely recognised source of sensor data is obtained from mobile smart devices. The ubiquitous nature of mobile smart devices such as smart phones, tablets, smart watch to name a few and the availability of cheap embedded sensors have completely revolutionised the smart city application dimensions.

## 3. SENSOR DATA FUSION

In this section we introduce sensor data fusion in the IoT domain. We also discuss its importance towards the IoT and where the techniques would fit into the IoT space.

As we discussed in earlier sections, IoT would produce substantial amount of data [55] that are less useful unless we are able to derive knowledge using them. We start our discussion by quoting some statements. The following statements strongly emphasis the necessity of sensor data fusion and filtering in IoT domain.

*"By 2020, wirelessly networked sensors in everything we own will form a new Web. But it will only be of value if the "terabyte torrent" of data it generates can be collected, analysed and interpreted [48]".*

*"Today, there are roughly 1.5 billion Internet-enabled PCs and over 1 billion Internet-enabled mobile smart phones. The present 'Internet of PCs' will move towards an 'Internet of Things' in which 50 to 100 billion devices will be connected to the Internet by 2020[68]".*

We see data fusion in the IoT environment as one of the most important challenges that need to be addressed to develop innovative services. In particular, in smart cities applications, when 50 to 100 billion devices start sensing [84], it would be essential to fuse, and reason about the data automatically and intelligently. Fusion is a broad term than can be interpreted in many ways. Hall and Llinas [27] have defined the sensor data fusion as a method of combining sensor data from multiple sensors to produce more accurate, more complete, and more dependable information that could not be possible to achieve through a single sensor. Nakamura et al. [52] have defined data fusion based on three key operations: complementary, redundant, and cooperative.

*Complementary* means putting bits and pieces of a large picture together. A single sensor cannot say much about the environment as it would be focused on measuring a single factor such as temperature. However, when we have data sensed through a number of different sensors, we can understand the environment in a much better way.

*Redundant* means that same environmental factor is sensed through different sensors. It helps to increase the accuracy of the data. For example, averaging the temperature value sensed by two sensors located in the same physical location would produce more accurate information compared to a single sensor. It also reduces the amount of data that need to be handled as it combines the two set of data streams together.

*Cooperative* operations combine the sensor data together to produce new knowledge. For example, reading RFID tags recorded in a supermarket can be used to identify the events such as shoplifting. Let's consider a scenario where RFID reader in a supermarket shelf detects that an item has been removed from a shelf. The RFID sensor in the counter does not see the object during payments. Later, the RFID sensor in the exit door detects the item that was removed from the shelf earlier. This sequence of actions can be simply inferred as a shoplifting event.

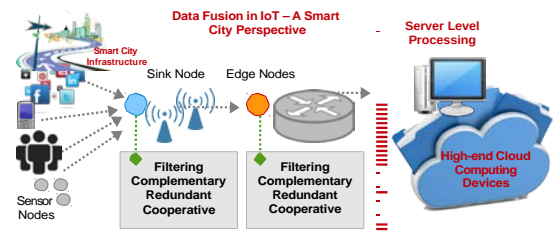

**Figure 1: Sensor Data processing**

A white paper published by Carnot Institutes [11] has listed data fusion and data filtering as two main challenges for the IoT and its applications such as smart cities. Data fusion is a data processing technique that associates, combines, aggregates, and integrates data from different sources. It helps to build knowledge about certain events and environments which is not possible using individual sensors separately. Data fusion also helps to build a context-awareness model that helps to understand situational context. The sensor data filtering stresses the requirement of filtering data to avoid large volumes of data transmission over the network.

The most basic sensor data fusion example that is used widely in smart phones is an e-compass. It uses a combination of 3D magnetometer and the accelerometer to provide compass functionality. Mainly, data fusion operations can be applied at two levels: cloud level and within the network level. As shown in Figure 1, sensor nodes, smart city infrastructure, edge node, sink nodes, and low level computational devices such as mo-bile phones belong to in-network sensor data processing. High-end computational devices such as servers belong to cloud level processing.

The cloud can help better understand the environment by performing complex sensor data fusion operations. Cloud level devices have access to unlimited resources and hence has the capability to apply complex data mining algorithms over the data generated by large number of lower level sensors. After understanding the environment, the cloud can generate actions that need to be taken appropriately.

In-network sensor data fusion is important to reduce the data transmission cost. As data transmission requires significant amount of energy, applying redundant fusion operation can reduce the data transmission. However, low-level nodes may not have the full view of the environment. Therefore, they may not be able to perform complex operations such as cooperative operations. The main responsibility of in-network sensor data fusion is to reduce the data transmission cost. The following rule defines how the data processing in each level should be conducted.

*L* = *CurrentLevel*;
if (*KnowledgeRequired* ≤ *KnowledgeAvailable*) ∧
 (*DataTransmissionCost* > *DataFusionCost*)
 then *ProcessAtTheCurrentLevel*(*L*)
  else *SendDataTo*(*L* + 1)

The ultimate goal of sensor data fusion is to understand the environment and act accordingly. This can be defined as a cycle as shown in Figure 2. We call it *Internet of Things Monitoring Cycle*. It has five steps: Collection, Collation, Evaluation, Decide, and Act. IoT monitoring cycle has been derived by combining the Intelligence Cycle [63] and the Boyd Control Loop [7]. The *Collection* step collects raw data from sensors and other IoT data sources (Social media, smart city infrastructure, mobile devices etc.). The *Collation* step analyse, compare and correlate the collected data. The *Evaluation* step fuses the data in order to understand and provide a full view of the environment. The *Decide* step decides the actions that need to be taken. The *Act* step simply applies the actions decided at the previous step. The *Act* step includes actuator control as well as sensor calibration and re-configuration.

Typically, the deployed IoT infrastructure in smart cities provide a means to monitor the environmental context [57, 59]. There is very little interest in the raw sensor data. The data that is of significant interest is information about interesting events that are happening in the specific area. In order to accomplish this task, IoT applications should be able to capture and reason about the events continuously. Therefore applying sensor data fusion techniques at the different levels of the IoT application chain is essential in order to detect relevant events.

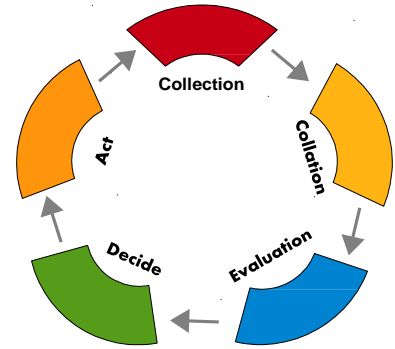

**Figure 2: Internet of Things Monitoring Cycle**

## 4. SENSOR DATA FUSION FOR SMART CITY APPLICATION

Data from citizens, systems, and general things flow through our cities thanks to the wide spread adoption of smart phones, sensor networks, social media and growing open release of datasets [3]. The data from Smart cities present a grand challenge to researchers and smart cities promoters, as we need to take advantage of these streams of information to build new services and define a clear return of investment for the benefit of the society [36].

The challenge in smart city is not to build a single generic model e.g. weather model based on temperature and humidity, complex models about noise pollution, traffic etc., but to combine all these together to build a good predictive contextually rich model. This model will help understand the dynamics of the society, and most importantly provide vital knowledge back to the citizens in order to enhance their quality of life.

A recent work from a group of researchers from MIT [65] demonstrate the potential of fusing data from disparate data sources in smart city to understand a city's attractiveness. The work focuses on cities in Spain and shows how the fusion of big data sets can provide insights into the way people visit cities. Such a correlation of data from a variety of data sources play a vital role in delivering services successfully in smart cities of the future.

In smart cities, ability to fuse sensor data enables context awareness which has a huge potential for IoT. Understanding the context of the city and its citizen can help develop and provide a new world of services based on what an individual user is doing, what the infrastructure [59] is doing, what nature is doing or all the above in various combinations [38]. The variety of services that can be developed is only limited to one's imagination. An example scenario could be a bridge experiencing a structural issue due to adverse environmental conditions can alert the city administrators and alert all cars travelling towards the bridge to stay away and seek alternative routes. For such a scenario to be feasible, it is important, smart city applications built on IoT have the ability to fuse data from diverse data sources to enable context-aware decision making and support.

## 5. EVALUATION FRAMEWORK

In this section, we present the framework that we use to evaluate different IoT sensor data fusion research efforts. The framework comprises the ten most significant features (parameters) related to sensor data processing in the IoT domain. Table 1 summarises the evaluation at the end of the Section 6.

### 5.1 Middleware Architecture Type

Middleware can be explained as a software layer that lies between the hardware and application layers. It provides the reusable functionalities that are required by the application to meet complex customer requirements. They are usually built to address the common issues in application development such as heterogeneity, interoperability, security, and dependability [34].

A traditional goal of middleware is to provide a set of programming abstractions to help software development where heterogeneous components need to be connected and communicated together (e.g: Internet of Things) [20]. However, programming abstraction comes at a cost. That means, when we use a middleware to connect sensors to applications, the performance will degrade due to additional overheads. If you manually connect application specific sensors to applications, they will perform much better. However, every time we develop a new application, we have to manually connect the sensors into the application where we will end up with repeated code. Compared to this repeated effort, using a middleware becomes a much better approach in term of cost and development time. Middleware systems are too general and are developed not for a single domain but for multiple domains. As a result, middleware may have functionalities that are not required by one application but that may be required in another application.

Todays' IoT applications demand more and more advanced and non-functional properties such as context-awareness and semantic interoperability. Middleware systems can bundles those functionalities together to be reused in many applications. We identify developing middleware as the right way to address the needs of IoT applications.

IoT (or sensor networks) middleware solutions can be mainly divided into two categories based on their installed location [31]: in-network schemes and server-side schemes. In-network middleware are usually developed using low level programming languages such as nesC [18, 75] and installed on each sensor node. Those middleware systems have more control on low level operation of the network such as network routing, energy consumption, etc. This layer is much closer to the hardware. However, it lacks the overall knowledge about the environment.

On the other hand, server-side middleware run in cloud computing environments. Those middleware collect data through gateways or sink nodes and are developed using high level programming languages such as C, Java etc. However, these middleware systems have less control over the sensor network operation. They are unable to control low level operations such as routing. However, they have more knowledge about the environment as they can analyse the sensor data received through different sensors. We have seen an emerging third category of middleware solutions, hybrid schemes, which combines both in-network and server side schemes. We believe that a hybrid middleware approach is best suited for the IoT domain as we can combine the best of both the in-network and cloud-based server schemes.

### 5.2 Context-awareness

The most widely used context information is location [22]. However, context in the IoT is much more broader than location. All the information about sensors can be considered as context information (e.g. capabilities of the sensors, related actuators, near by sensors, etc.). With the recent advancement of the IoT, context-awareness has become an essential part of the IoT applications. Context-awareness is no more limited to mobile applications. Currently, the largest context information consumers are mobile devices and their applications. A research effort called *mSense* [41] has introduced a middleware solution to manage context information. *mSense* separates context-awareness management functionalities into a separate layer. The IoT domain also requires such separation to make application development much easier and faster.

Chantzara and Anagnostou [13] have identified four common stages in context-aware application life cycle as context sensing, context processing, context dissemination, and context usage. This life-cycle has been enhanced by [32]. Combining sensor data from multiple sensors helps to understand context information much more accurately. Better understanding will contribute towards intelligent fusion and complex event detection.

The Cluster of European Research Projects (CERP-IoT) has also mentioned context awareness (location-aware, environment aware) as a key characteristic of objects in the IoT space [68]. Identifying the context information such as geographical location, sensor capabil-ities, near-by sensors, related actuators and supported data formats would help to built a context model for each sensor that can be used to increase the autonomous interaction among sensors. Nagy et al. [51] have defined a term called *Global Understanding* in related to context-awareness. It means that sensor *'A'* can understand the properties and capabilities of sensor *'B'* and vice versa. This can only be achieved through semantic technologies and context awareness techniques.

A research focused on *smart objects* [40] has defined three types of context-awareness: activity-aware, policy-aware, and process-aware. Activity-aware means the ability to understand the activity and the usage of a specific sensor. Policy-aware acts as a domain knowledge repository where it consists of rules. For example, policy-aware can identify the health and safety conditions of the user via policy knowledge and act accordingly. Process-aware is the ability to detect the current processes carried out by the user and the surrounded objects. An ideal IoT application should be able to provide additional assistance to users to carry out their work as mentioned above.

Abowd and Mynatt [1] have identified *5Ws* (who, what, where, when, why) as the minimum set of context information that need to be handled in a pervasive computing environment. This stays true in the IoT space as well. Context information can be divided into three categories: user context, computing (system) context, and physical (Environmental) context [61]. User context means the knowledge about the user (e.g. age, gender, likes, dislikes, etc.). Computing context means the knowledge about the software and hardware used by users (e.g. operating system, hardware capacity, software applications, etc.). Physical context means the knowledge of the environment such as location, temperature, light, etc.

Issarny et al. [34] have distinguished three types of context sensitivity: context-specific systems, context-dependent systems, and context-adaptive systems. Applications that can work only in one context are called context-specific. Context-dependent applications need to be configured at the beginning of the application for each context. Context-adaptive systems can change their behaviour dynamically during runtime when context changes. IoT applications demand the context-adaptive behaviour to make the IoT vision a reality.

In order to build a comprehensive context model using context information, it is necessary to acquire context data through many different data sources. A single source would not be able to provide all necessary information that can be used to understand the context accurately. Therefore, combining the context information retrieved through multiple sources is essential but challenging [44].

### 5.3 Semantic Interaction

The IoT can be considered as an application domain where semantic web technologies can be used to enhance its functionalities significantly [30]. The IoT promises to connect the billions of things around us together. It is not feasible to manually connect by hard-wiring things and applications. Automating these tasks will definitely need the help of semantic technologies. Research conducted on semantic sensor web [15] has identified several challenges that need to be addressed by semantic technologies. For example, sensor configuration, context identification, complex sensor data querying, event detection and monitoring, and sensor data fusion are some of the tasks that can be enhanced using semantics. Annotating sensors, sensor data, and program components will increase the ability of interaction without explicit programming commands. Furthermore, annotations will also increase the retrievability of sensor data. More sophisticated queries can be processed over the semantic annotated data.

### 5.4 Dynamic Configuration

Dynamic configuration can be interpreted at two levels: a software level dynamic configuration and a hardware level dynamic configuration. Dynamic hardware configuration stresses the adaptability of a system. IoT comprises tiny sensing devices (things) which are prone to fail frequently. Therefore, a network built by these devices is unreliable and should be able to change, configure and adapt itself to the environment dynamically. Furthermore, things may need to change their configuration as a result of the decisions made by the cloud-based server as a part of the actuation control.

For example, lets consider a things (sensor node) *S* that is capable of sensing light, temperature and humidity. It is physically located in area *L*. Currently, sensor node *S* measures only temperature as it is the expected requirement of the server level software to make the decisions. Later, the server may require to know the light level of area *L*. The sensor node *S* needs to be configured to measure not only temperature, but also light level as well. This new configuration setting needs to be pushed to the sensor node from the cloud server. Figure 3 presents an example of a dynamic reconfiguration for wireless sensors nodes deployments. According to our survey, this functionality is lacking among most of the current research efforts.

Furthermore, software level can also support dynamic configuration capabilities. For example, software components described in semantic technologies can be combined together to create complex data fusion operations. Rather than combining these components at development time, runtime configuration can add more adaptability to the system.

The complex data fusion operations should be built by reusing the software components at runtime based on the user requirements. For example, Figure 4 shows how a system can dynamically configure the components into a work flow in order to detect events and act accordingly.

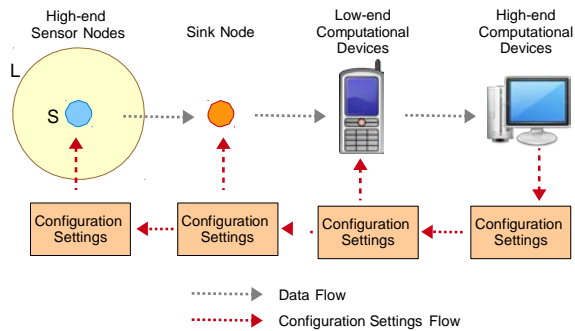

**Figure 3: Dynamic Sensor Network Configuration**

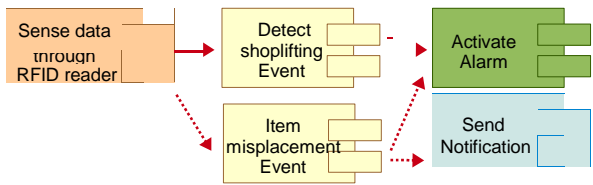

**Figure 4: Software Dynamic Configuration**

### 5.5 Fusion Complexity

Querying data from *things* is one of the common data fusion operations in the IoT domain [8, 24, 46, 80]. The level of complexity supported by the query may differ from the query language implementation. Sometimes, semantic technologies such as SPARQL [47], are used to query sensor data.

Another common data fusion approach is event detection. Events can be recognised by identifying and correlating sequences of action that occurred in the environment. Lets consider two sensors *A* and *B*, where *A* measures temperature and *B* is a camera. In an ideal system, users should be able to pose queries to retrieve the video feed of a room where temperature is higher than 35°C. In order to answer such a request, the system should be able to combine both sensors *A* and *B* together. Another example query would be 'identify the best place to store a sculpture in a museum based on the sculpture specification'. A number of data fusion operation need to be used to answer such queries. Concretely, a query may need to be generated by using optimum humidity level, temperature and other parameters.

This kind of combining needs to be supported by semantic technologies. Song et al. [67] have provided a full description on how to accomplish such tasks by using semantic technologies. Zafeiropoulos et al. [67] have described all the elements such as sensors and programming modules using semantics, where complex combinations are possible.

### 5.6 Actuation Management

According to our evaluation, the majority of research efforts have left out the functionality of actuation management from their proposed solutions. We presented the IoT monitoring cycle in Section 3. This cycle stresses the importance of the *act* step. Sensors sense data and transmit it to servers. Severs then do the processing and take the decision on how to handle the situation based on the gathered knowledge and previous experience. Then actions need to be taken. Action can be a change in sensor configuration or to conduct a specific task using a connected actuator. For example, actuation could increase the humidity by spraying more water into the air. According to the context, the most appropriate actions needs to be taken and managed by an ideal IoT software system in an efficient manner.

### 5.7 Type of Processing

Data processing in IoT can be done in two ways: in-network processing and cloud level processing. Sensors are prone to produce faulty data due to technical issues. Furthermore, sensors produce redundant data that wastes the energy if they are transmitted. Therefore, data filtering is critical to save energy. In-network processing mechanisms can be used to address these issues. In-network sensor data processing however has limitations, because in-network devices such as sensor nodes and mobile phones perceive only limited knowledge about the environment (local context) [54]. Therefore, in-network data processing cannot make high level decision where overall knowledge is required.

Cloud-based processing should be used to address the above problem. Cloud servers receive all the data collected through a variety of different sensors. These data increase the knowledge about the environment, so the servers can take decisions by considering overall knowledge (global context). Furthermore, cloud-based server devices have more sophisticated hardware power to process and understand large amounts of data compared to in-network hardware. Server level sensor data processing techniques are used to fuse data in many ways according to user queries. It can also understand interesting events that occur in a sensor network.

It is clear that both types of processing haves their unique contribution towards sensor data fusion in the IoT domain. Therefore, the ideal way to process sensor data is to use a hybrid approach where both in-network and cloud-level sensor data processing techniques are employed.

## 5.8 Cross Domain Portability

Cross domain portability stresses the ability of applying a proposed solution on different domains. Most of the proposed solutions are narrowly focused on one domain. We believe it is ideal to implement a solution addressing more than one domain in order to prove the cross domain portability. At the same time, it is critical to differentiate the domain specific and domain independent components of a solution. This increases the ability to apply a solution in different domains. A clear differentiation will enable rapid and easy expansion.

## 5.9 Implementation

Implementation is critical in order to prove a concept. Challenges that cannot been seen in theoretical process can be clearly seen in a practical implementation. Implementation allows the identification of the practical and technical difficulties and challenges that arise during the implementation process. The majority of proposed solutions are practically implemented. Proper implementation should be followed by a rigorous performance evaluation procedure. The choice of programming model, platform and languages significantly impact on the future development and scalability. Making the programming code *open source* is a one approach that can ensure the rapid future development and avoid repetitive work among researchers.

## 5.10 Performance Evaluation

Performance of a system becomes critical when the system becomes larger and larger. In the IoT, we expect to connect millions and billions of sensors together. Therefore, performance evaluation is critical to understand and verify how the system would work in a real world deployment. It also allows to optimise the solution based on the performance evaluation results. Unfortunately, most of the proposed solutions in the IoT domain have not conducted a performance evaluation procedure which makes hard to decide the applicability of the proposed solutions in real world.

Performance evaluation remains an open issue and a challenge that needs to be achieved by researcher in the IoT domain. Performance evaluation can be categorised into two distinct areas: software and hardware. Parameters such as energy, response time, data fusion capability, and number of supported sensors, need to be evaluated.

## 6. SENSOR DATA FUSION APPROACHES - STATE-OF-THE-ART

In this section, we discuss some of the solutions proposed by different researchers. We highlight the significances of each project in brief. At the end, a summary of the evaluation is presented in Table 1. It is to be noted, a number of solutions in sensor data fusion have been addressed within the wireless sensor network research. These solutions are completely applicable with the IoT domain.

Jara et. al. [36] have applied sensor data fusion to understand human behaviours in smart cities. Their work analyses data obtained from the European project Smart Santander. The work demonstrates how ubiquitously available data such as traffic flows and temperature can be correlated to understand and model the influence of temperature on traffic flow. The work considers the Poisson model and shows that the Poisson distribution model is not always valid.

Sobolevsky et al. [65] have applied sensor data fusion to estimate the attractiveness of smart cities for visitors. The work focuses exclusively on cities in Spain. To arrive at attractiveness they fuse sensor data from three data source namely credit and debit card transactions carried out by visitors, 3.5 million photos and videos taken in spain and posted to Flickr and 700,000 geo-tagged tweets. The attractiveness of a city for the purpose of the city was defined as the total number of tweets, pictures and card transactions that took place within it. The work produces some interesting results and demonstrated how fusion of sensor data sets (big data sets) can provide insights into how people use cities. In general, the work identified bigger cities attract large number of visitors. However, there were also some exceptions that deviate from the above assumption. For example certain cities such as Malaga had high level of visitors but the least number of Flickr activity. This is due to the fact, these cities are considered as retirement locations and the category of visitors tends to less use social media such as Flickr. This work is an excellent demonstration of how data fusion in smart cities can help create innovative services delivering value back to its citizens and smart city developers.

Antonelli et al. [3] present city sensor fusion, a big data platform that collects, aggregates, analyses, semantically enriches and offers visual analytics from data flows in smart cities. The work focuses on using sensor data fusion to detect city scale events such as event lasting days, number of visitors attracted, venues that attracted significant interest etc. The platform fuses data from different types of data sources ranging from social media to mobile phones to sensors such as Traffic flow, weather and pollution.

Soldatos et al., [66] propose OpenIoT a first-of-kind open source IoT platform enabling the semantic interoperability of IoT services in the cloud. OpenIoT promotes interoperability among IoT silos right from the sensor to the cloud services. OpenIoT is built upon semantic web standards such as W3C Semantic Sensor Networks (SSN) ontology, which provides a common standards-based model for representing physical and virtual sensors, RDF to store, index and retrieve data, and supports virtually any IoT protocols such as CoAP, 6LoWPAN etc. OpenIoT includes also sensor middleware and sensor data fusion capability at the things and at the cloud. OpenIoT eases the collection of data from virtually any sensor, while at the same time ensuring they are embedded with proper semantic annotation. Furthermore, it offers a wide range of Do-it-yourself visual tools that enable the development and deployment of IoT services and applications with almost zero programming. Another key feature of OpenIoT is its support for mobile sensors and thereby enabling support for an emerging wave of mobile crowd sensing applications. The OpenIoT platform is a blueprint architecture to develop semantically interoperable smart city solutions with support for complex sensor data fusion algorithms.

Zanella et al., [83] offers a survey of available techniques, architecture, and protocols for a urban IoT which are used to achieve the Smart City" vision. The paper describes characteristics of an urban IoT and overviews some services related to Smart City. The technical solution proposed in this paper have been used in Padova (Italy) Smart City project [12]. The Padova project employs IPv4 and IPv6 at th network layers and uses a wireless sensor network gateway to collect data from deployed sensor network infrastructure [59]. Theodoridis et al., [72] illustrates challenges, socioeconomic chances and vital findings from the European smat city project Smart Santander. The paper surveys a Logical 3-tier node and 3-plane architecture and highlights various use cases that employ sensor data fusion in smart cities including Outdoor parking management, precision irrigation and home garden monitoring. Lin et al., [37] presents an information framework which encompasses the complete urban information system for building a Smart City by using the Internet of Things. The paper use a Noise Mapping in Smart Cities case study to demonstrate the architecture.

Da Rocha et al. [17] have focused on developing semantic middleware for wireless sensor networks using low level programming (i.e using NesC, a extension to the C programming language used for embedded programming). The approach is based on a rule-based reasoning engine using ontologies. The research addresses the Structural Health Monitoring (SHM) application domain. Research justifies the reason of choosing wireless sensor networks over wired sensor networks by pointing out the fact that wired sensor networks are time consuming to deploy, very expensive and hard to reconfigure [17].

Semantic sensor networks in SHM domain enable the usage of semantic information towards monitoring and handling the environment. The research incorporates semantic features at the middleware level using a low level programming approach. The middleware has been implemented using the NesC language in Mica Motes [16] that runs the TinyOS [74] operating system. The reasoning engine Pellet [14] is integrated in this middleware. New behaviours can be added into the application by adding new rules. All the communication between the nodes are done by using a XML format called *TinyXML* [73]. Knowledge is stored and processed using OWL. Da Rocha et al. [17] have developed ontologies related to the domain and other services. Application driven, device driven and network driven concepts are defined in the ontology. Ontologies help to share information such as power remaining on a sensor, capabilities of the sensors and so on.

The middleware proposed by Da Rocha et al. [17] intelligently shares information between different sensors based on semantic knowledge. For example, two sensors in the same area should not share their information if those sensors are measuring two different aspects of the environment; for example light and corrosion. However, if the two sensor measurements complement each other, such as humidity and corrosion, then the sensors should share their measurements and do the reasoning by combining both measurements. When many sensors measure the same aspect, few of the sensors can switch themselves off intelligently to save energy resources.

Zafeiropoulos et al. [81, 82] have presented an architecture to address the issues such as data aggregation, data management, and querying. The semantic technologies are used to extract meaningful information from the raw sensor data. Aggregation of data contains less value unless they are interpreted accurately. The interpretation is essential in order to detect interesting events in sensor networks. Zafeiropoulos et al. [81, 82] correctly argue that this event detection should be supported by data gathered through heterogeneous data sources. The semantic technologies that support such operations are content description languages, query languages, and annotation frameworks. The proposed architecture comprises three layers: data layer, processing layer, and semantic layer. The data layer is responsible for collecting data from sensors using event-based or polling-based mechanisms. The processing layer converts those raw data into XML files. In the semantic layer maps the XML data into a semantic model where the XML messages are stored in the form of class individuals. This conversion is done by XML mapping rules. Another set of rules called *semantic rules* are used to detect events. As a result of these conversions, a system can query and reason the sensor data using semantic query languages which provide enriched capabilities.

The project *Hydra* [21] addresses the needs of healthcare, home automation and agriculture domains. It provides an architecture to connect sensor devices together to detect events. The *Hydra* middleware is based on a Service Oriented Architecture (SOA) and a Model Driven Architecture (MDA). The core architecture of *Hydra* comprises a number of different managers, such as network, discovery, ontology, event, storage, and context managers. Each of these managers are divided into a number of layers. For example, the context manager comprises four layers. Context data acquisition, context management, context awareness, and context reasoning and interpretation. The *Hydra* middleware does not differentiate the domain specific and domain interpreted components in its architecture, which makes it hard to extend the domain into other domains. *Hydra* encapsulate sensors into web services and the devices are described using semantics where it enables semantic interoperability among sensors. However, data is not annotated using semantics.

Lee et al. [42] have proposed a hybrid middleware which comprises an in-network middleware and a server-side middleware. The in-network middleware has the capability to deal with operations such as energy efficient data transmission. The server-side middleware handles the context-aware stream processing, event detection and querying. The main focus is given to the in-network middleware. Therefore, event detection and data fusion capabilities are very limited. The in-network middleware has the intelligent capability to identify incomplete and false data values.

Bruckner et al.[9] have proposed a framework to process audio and video sensor data in a semantic manner. The proposed system architecture comprises seven layers. The bottom layers which are closed to the sensor nodes do the image and audio processing and convert the raw data into *Low Level Symbols (LLS)*. Then data fusion mechanisms are used to convert those symbols into *High Level Symbols (HLS)*. Patterns and events can be recognised using these symbols. The implementation has been deployed in an airport domain where the system is capable of identifying events such as unattended luggage or gunfire. The entire architecture is narrowly focused on video and audio sensor data processing.

*Semantic Sensors* (*SS*) [33] network middleware connects a variety of sensors to applications. The objective of the middleware is to develop a sensor network where developers need not to be aware of the device type of each sensor node. *SS*t can identify the location and the relationship among the sensors. The evaluation of the middleware has been done in a lab environment by attaching sensors to daily use items such as bottles and books. Logical expressions are used to store information about each object and their relationships. Very primitive events are possible to recognise by the system. For example, the system can answer simple queries such as identify the state of the object (i.e moving or not) or recognise the other objects near by. The implementation is done using low level programming languages such as nesC.

*Semantic Web Architecture for Sensor Networks* (*SWASN*) [29] is a server-side middleware that uses semantic web technologies to enrich sensor data processing. This project has proposed a four layer architecture: sensor networks data sources layer, ontology layer, semantic web processing layer, and application layer. *SWASN* is capable of connecting multiple sensor networks together. To achieve this challenge, *SWASN* uses a separate local ontology for each sensor network to map sensor data to a common global RDF data model. *SWASN* provides sophisticated querying features using SPARQL [47]. The system is focused on building fire emergency domain.

*u-Greenhouse* [31] is a context-aware middleware that proposed to process data collected through sensors in a greenhouse environment by applying wireless sensor network technologies. This middleware provides the functionalities of data filtering, event processing, context-aware processing and integration of heterogeneous sensors. The system architecture consists of three parts: sensor network interface, data process, and application service interface. The approach is to develop a hybrid middleware that consists of in-network data processing middleware that are installed on each node and a server-side data processing middleware. The *u-Greenhouse* architecture comprises three layers: the physical layer (Sensor node and gateways), the middleware layer and the application layer. Semantic capabilities are provided using and context-aware ontology. The system is capable of recognising simple events in greenhouse environment that can trigger actions. *u-Greenhouse* solution is narrowly focused on greenhouse domain.

Siguenza et al. [64] combine *states chart* technology and semantic technology to annotate and process sensor data. The objective is to derive high level information from raw sensor data. W3C State Chart eXtensible Markup Language (SCXML) is used to implement the system. The sensor data are enriched using RDF semantics and stored in an SCXML data model. The possible situations are defined as states such as *adverseWeather*. The conditions related to the *adverseWeather* state need to be fulfilled in order to infer the current state as *adverseWeather*.

*HARMONI* [28] is a context-aware system for the healthcare domain. This project has gone beyond the objective of identifying events using sensor data fusion. Homed et al. [28] have used their framework to reduce the amount of data transmission significantly. A mobile device that is capable of filtering data is deployed in the patient's room. This device is able to monitor the events according to the specifications defined in the filters. For example, doctors may need not to know all the behaviour of a patient. Doctors are only interested to know when a patient shows any unusual behaviour

(e.g. very high heart rate). Therefore, it is not necessary to transfer all the data sensed by the sensors to the back-end server. Instead, a mobile device in the patient's room can filter the sensed data and transfer only the relevant data intelligently to the server based on the filter definition [28]. These filters need be changed according to the context. For example, heart rate may need to be monitored based on the context. When the patient is doing exercises, it is natural that heart rate goes up. Therefore, the filters should be able to alter the filter definitions based on the context.

Terziyan et al. [70] have proposed the *UbiRoad* middleware that uses semantic and agent technologies. Their focus is on the smart road and traffic control domain. A sample scenario has presented to convey the ultimate objective. *UbiRoad* addresses four main challenges: interoperability, flexible coordination, self-management, trust and reputation. They have also identified context-aware sensor data fusion as a secondary challenge. The proposed solution is based on two other projects *SmartResource* [69] and *UBIWARE* [39]. *UbiRoad* combines different ontologies to incorporate different concepts into the system such as device ontology, context ontology, data ontology, and domain ontology. For example, device ontology is used to recognise different devices in the sensor network and context ontology is used to understand the traffic control domain.

Phuoc and Hauswirth [60] have proposed the concept of combining link data towards sensor data mashups. The system acquires sensor data through wrappers and passes them to the upper layer for fusion operations. The data fusion comprises many operations such as data filtering, data alignment, association, correlation, pattern detection and classification. Fusion operations can be composed together to produce high-level filters. The acquired sensor data are stored in RDF models. Therefore, SPARQL is used to query the data. Each individual sensor in considered as sensor component. Combinations of sensor components are defined as a sensing system. Sensor systems and fusion operation can be combined together to build complex work flows. An Ajax [23] based graphical user interface is provided to build those work flows. The approach is more focused towards utilising link data concepts.

Gyllstrom et al. [26] have proposed a complex event processing system over data streams called *SASE*. *SASE* is narrowly focused on the RFID sensors domain. A high-level SQL like language has been defined to support user queries. The system is capable of identifying events such as shoplifting or inventory misplacement. Users need to syntactically define the query, and the system can process the query against the data stores. Some data fusion operations such as anomaly filtering, temporal smoothing and duplication reduction are provided by *SASE*.

Liu and Zhao [43] have identified that most of the efforts on sensing systems today are domain specific with very little re-usability. To solve this problem, they have proposed a open architecture which is enriched with semantics. XML data formats are used in the system to store data. Service components are the main building block in the system. Each service is designed to take inputs, do some processing and give the output back. Services are designed in such a way that multiple services can be combined together to build a complex service. This run-time combining process is possible due to semantic descriptions. This programming model allows the user to query the sensor data and events in abstract ways without dealing with raw sensor data.

*SEMbySEM* [10] is a sensor management framework that focuses on isolating technical related challenges from the applications layer by using a facade layer in-between. The facade layer transforms the sensor data into semantically enriched information. The proposed architecture comprises three layers: facade, core, and visualisation. The core layer does the reasoning and inferring. An ontological semantic model is used to store the concepts, rules and data.

*Intelligent Event Processing Agent* (*iEPA*) [19] is an approach that combines complex event processing and multi agent systems. The research is focused on traffic management domain. A rule based system is employed to identify the events. Data fusion operations such as filter, split, aggregate, and transform are used to infer events. Events are defined in a language called *Espers continuous Query Language* (*EQL*).

Izumi et al. [35] have proposed a knowledge filtering scheme for the health care support domain. Their system comprises a number of different agents, such as a data stream mining agent, a inference agent, and a knowledge base agent. A multi agent architecture is used to built the system, and an ontology scheme is used to store data where SPARQL queries are used for data filtering. Knowledge gathered using sensors is filtered based on four different perspectives: person based filtering, access policy based filtering, location based filtering, and time based filtering.

Teymourian et al. [71] present a conceptual approach to address the problem of *Semantic Event Processing* (*SCEP*). *SCEP* combines event processing technologies and semantic technologies. This research effort is not directly related to sensor data fusion. However, the techniques used in this area can be combined with sensor data stream processing in order to detect events in the IoT environment.

The *Sensor Web Agent Platform* (*SWAP*) framework [50] comprises three layers: sensor layer, knowledge layer, and application layer. A multi agent technology and web services technologies are employed to built the system. Each layer consists of a number of agents that are capable of doing specific tasks. The implementation is focused on a fire detection domain. The number of different agents can be combined together to answer or detect complex situations such as wildfire.

## 7. EVALUATION OF SENSOR DATA FUSION APPROACHES

The Table 1 classifies the difference sensor data fusion efforts based on the evaluation framework we presented in section 5.

The parameters used to evaluate each feature of research efforts can be explained as follows. In depth discussion on each feature is conducted in the Section 5.

- *Architecture Type:* This feature evaluates whether the proposed solution is proposed as a middleware (M) or an Application system (A). Application systems are narrowly focused on one specific domain while middleware solutions possesses more domain expandability and domain independence.

- *Context-awareness:* This feature evaluates whether the proposed solution possesses context-awareness capabilities or not.

- *Semantic Interaction:* This feature is evaluated using four categories: High (H), Moderate (M), Low (L), and none (×).
  – High (H) - Both data and program components are annotated using semantic technologies. Semantic reasoning mechanisms are employed.
  – Moderate (M) - Either data elements or program components are enriched using semantics technologies, but not both.
  – Low (L) - No semantic technologies are used. However, solutions are enriched with limited semantic capabilities using different techniques such as rules [9], symbols [33], etc.
  – None (×) - No semantic interactions posed by the approach.

- *Dynamic Configuration:* This feature is evaluated using four categories: High (H), Moderate (M), Low (L), and none (×).
  – High (H) - Sensor hardware and software components are dynamically configured based on the environment. The solution possesses automated configuration of filtering, fusion and reasoning mechanism, according to the problems at hand.
  – Moderate (M) - Poses very limited dynamic hardware configuration such as switch on/off sensors.
  – Low (L) - Poses software level limited dynamic composition and configuration capabilities.
  – None (×) - No software or hardware components are dynamically configured.

- *Fusion Complexity:* This feature is evaluated using three categories: High (H), Moderate (M), and Low (L).
  – High (H) - Capable of answering complex user queries. Program components can be combined together to produce complex results.
  – Moderate (M) - Capable of answering moderately complex user queries. Develop complex fusion mechanism by combining simple fusion components is not possible.
  – Low (L) - Limited fusion techniques such as data filtering is possible.

- *Actuation Management:* Does the solution possesses actuation management capabilities.

- *Type of Processing:* Is the data fusion approach Centralised (C) or Decentralised (D).

- *Cross Domain Portability:* Number of domains that the proposed solution is applied.

- *Implementation:* This feature tells that whether researchers have practically implemented the proposed solution or if it is a theoretical approach only.

- *Performance Evaluation:* This feature evaluates whether each research effort has conducted a performance evaluation procedure on their proposed system or not.

## 8. CONCLUSION

In this article, we first highlighted the importance of sensor data fusion in IoT application such as smart cities applications. We examined a number of different sensor data fusion research efforts related to IoT with particular focus on smart cities application domain.

We developed a evaluation framework by carefully selecting ten different metrics. We believe these ten metrics are open challenges in the field. Some of these challenges are addressed by the researchers significantly and some are in its infancy. One of the major goals of this article is to highlight the opportunities for improvements and research gaps in the field.

Based on surveyed approaches, context-awareness in IoT more specifically within the smart city domain is gaining importance but still in its infancy. A lot of focus on context awareness is towards a particular application while to realise the true IoT-enabled smart cities vision, a broader non-domain focus will have to be pursued. Furthermore, dynamic configuration of things is also not addressed by most of the proposed solutions. Similarly, actuation management is the least addressed feature among all. We believe actuation management is important as it plays a significant role in the IoT monitoring and feedback cycle. Further, performance evaluation techniques employed by most of the researchers to evaluate their proposed approaches are limited. Performance evaluation is extremely important as we are expecting these solutions to incorporate billions of sensor devices. Finally, cross domain portability is also addressed poorly. The majority of the efforts are based on a single domain. It is hoped that future efforts will aim to address these research gaps.

| Research Efforts | Architecture Type | Context-awareness | Semantic Interaction | Dynamic Configuration | Fusion Complexity | Actuation Management | Type of Processing | Cross Domain Portability | Implementation | Performance Evaluation | Year |
|---|---|---|---|---|---|---|---|---|---|---|---|
| Gibbons et al. [24] | A | × | × | L | L | × | D | 3 | ./ | × | 2003 |
| Liu & Zhao [43] | A | × | H | M | H | × | C | 1 | ./ | × | 2005 |
| Whitehouse et al. [77] | M | × | M | L | H | × | C | 1 | ./ | × | 2006 |
| Lewis et al. [49] | M | × | H | M | M | × | C | 1 | ./ | × | 2006 |
| Moodley et al. [50] | M | ./ | L | × | L | × | C | 1 | ./ | × | 2006 |
| Moodley & Simonis [50] | M | × | M | M | H | × | C | 1 | ./ | × | 2006 |
| Bouillet et al. [6] | M | ./ | H | L | H | ./ | C | 1 | ./ | ./ | 2007 |
| Brenna et al. [8] | A | ./ | × | × | M | × | C | 1 | ./ | × | 2007 |
| Gyllstrom et al. [26] | A | × | × | × | M | × | C | 1 | ./ | × | 2007 |
| Noguchi et al. [26] | M | × | M | H | M | × | C | 1 | ./ | × | 2007 |
| Zafeiropoulos et al. [81] | M | × | H | L | M | × | C | 1 | ./ | ./ | 2008 |
| Sheth et al. [62] | A | × | H | L | M | × | C | 2 | ./ | × | 2008 |
| Bruckner et al. [9] | A | ./ | L | × | M | × | D | 1 | ./ | × | 2008 |
| Huang et al. [29] | M | ./ | H | M | M | × | C | 1 | ./ | × | 2008 |
| Wood et al. [78] | A | ./ | L | M | L | × | D | 1 | ./ | × | 2008 |
| Homed et al. [28] | A | ./ | L | M | H | ./ | D | 1 | ./ | ./ | 2008 |
| Ni et al. [53] | A | ./ | H | × | M | × | D | 1 | × | × | 2009 |
| Da Rocha et al. [17] | M | ./ | M | L | M | × | D | 1 | ./ | × | 2009 |
| Phuoc & Hauswirth [60] | M | × | H | L | H | × | C | 1 | ./ | × | 2009 |
| Teymourian et al. [71] | A | ./ | L | L | L | × | C | 2 | × | × | 2009 |
| Brunner et al. [10] | M | × | H | × | M | × | C | 1 | ./ | × | 2009 |
| Eisenhauer et al. [21] | M | ./ | H | M | H | × | C | 2 | ./ | × | 2009 |
| Lee et al. [42] | M | × | L | × | L | × | D | 1 | ./ | ./ | 2010 |
| Siguenza et al. [64] | A | × | M | × | L | × | C | 1 | ./ | × | 2010 |
| Izumi et al. [35] | A | ./ | M | × | M | × | C | 1 | ./ | ./ | 2010 |
| Terziyan et al. [70] | M | ./ | M | L | M | × | C | 1 | ./ | × | 2010 |
| Hwang et al. [31] | M | ./ | L | M | L | ./ | D | 1 | ./ | ./ | 2011 |
| Dunkel [19] | A | × | L | M | H | ./ | D | 1 | ./ | × | 2011 |
| Zanella et al. [83] Theodoridis et al. [72] Lin et al. [37] | M | × | L | L | M | L | C | 1 | ./ | ./ | 2014 |
| Jara et al. [36] | A | ./ | L | L | H | ./ | C | 1 | ./ | ./ | 2014 |
| Sobolevsky et al. [65] | A | ./ | L | L | H | ./ | C | 1 | ./ | ./ | 2015 |
| Antonelli et al. [3] | A | ./ | H | M | H | ./ | C | 1 | ./ | ./ | 2014 |
| Soldatos et al. [66] | M | ./ | H | H | H | ./ | D | many(5) | ./ | ./ | 2014 |

**Table 1: Taxonomy of Sensor Data Fusion Research Efforts**